\documentstyle[]{article}
\voffset= - 1.5in
\catcode`\@=11
\def\marginnote#1{}

\newcount\hour
\newcount\minute
\newtoks\amorpm
\hour=\time\divide\hour by60
\minute=\time{\multiply\hour by60 \global\advance\minute by-\hour}
\edef\standardtime{{\ifnum\hour<12 \global\amorpm={am}%
        \else\global\amorpm={pm}\advance\hour by-12 \fi
        \ifnum\hour=0 \hour=12 \fi
        \number\hour:\ifnum\minute<10 0\fi\number\minute\the\amorpm}}
\edef\militarytime{\number\hour:\ifnum\minute<10 0\fi\number\minute}

\def\draftlabel#1{{\@bsphack\if@filesw {\let\thepage\relax
   \xdef\@gtempa{\write\@auxout{\string
      \newlabel{#1}{{\@currentlabel}{\thepage}}}}}\@gtempa
   \if@nobreak \ifvmode\nobreak\fi\fi\fi\@esphack}
        \gdef\@eqnlabel{#1}}
\def\@eqnlabel{}
\def\@vacuum{}
\def\draftmarginnote#1{\marginpar{\raggedright\scriptsize\tt#1}}
\def\draft{\oddsidemargin -.5truein
        \def\@oddfoot{\sl preliminary draft \hfil
        \rm\thepage\hfil\sl\today\quad\militarytime}
        \let\@evenfoot\@oddfoot \overfullrule 3pt
        \let\label=\draftlabel
        \let\marginnote=\draftmarginnote
   \def\@eqnnum{(\theequation)\rlap{\kern\marginparsep\tt\@eqnlabel}%
\global\let\@eqnlabel\@vacuum}  }

\def\d{\partial}

\def\bea{\begin{eqnarray}}
\def\eea{\end{eqnarray}}
\def\nn{\nonumber}
\def\beq{\begin{equation}}
\def\eeq{\end{equation}}
\def\ba{\beq\new\begin{array}{c}}
\def\ea{\end{array}\eeq}
\def\be{\ba}
\def\ee{\ea}
\def\stackreb#1#2{\mathrel{\mathop{#2}\limits_{#1}}}
\def\Tr{{\rm Tr}}

\makeatletter
\newdimen\normalarrayskip              
\newdimen\minarrayskip                 
\normalarrayskip\baselineskip
\minarrayskip\jot
\newif\ifold             \oldtrue            \def\new{\oldfalse}
\def\arraymode{\ifold\relax\else\displaystyle\fi} 
\def\eqnumphantom{\phantom{(\theequation)}}     
\def\@arrayskip{\ifold\baselineskip\z@\lineskip\z@
     \else
     \baselineskip\minarrayskip\lineskip2\minarrayskip\fi}
\def\@arrayclassz{\ifcase \@lastchclass \@acolampacol \or
\@ampacol \or \or \or \@addamp \or
   \@acolampacol \or \@firstampfalse \@acol \fi
\edef\@preamble{\@preamble
  \ifcase \@chnum
     \hfil$\relax\arraymode\@sharp$\hfil
     \or $\relax\arraymode\@sharp$\hfil
     \or \hfil$\relax\arraymode\@sharp$\fi}}
\def\@array[#1]#2{\setbox\@arstrutbox=\hbox{\vrule
     height\arraystretch \ht\strutbox
     depth\arraystretch \dp\strutbox
     width\z@}\@mkpream{#2}\edef\@preamble{\halign
\noexpand\@halignto
\bgroup \tabskip\z@ \@arstrut \@preamble \tabskip\z@ \cr}%
\let\@startpbox\@@startpbox \let\@endpbox\@@endpbox
  \if #1t\vtop \else \if#1b\vbox \else \vcenter \fi\fi
  \bgroup \let\par\relax
  \let\@sharp##\let\protect\relax
  \@arrayskip\@preamble}
\def\eqnarray{\stepcounter{equation}%
              \let\@currentlabel=\theequation
              \global\@eqnswtrue
              \global\@eqcnt\z@
              \tabskip\@centering
              \let\\=\@eqncr
              \\%
 \halign to \displaywidth\bgroup
    \eqnumphantom\@eqnsel\hskip\@centering
    $\displaystyle \tabskip\z@ {##}$%
    \global\@eqcnt\@ne \hskip 2\arraycolsep
         $\displaystyle\arraymode{##}$\hfil
    \global\@eqcnt\tw@ \hskip 2\arraycolsep
         $\displaystyle\tabskip\z@{##}$\hfil
         \tabskip\@centering
    &{##}\tabskip\z@\cr}
\begingroup\ifx\undefined\newsymbol \else\def\input#1 {\endgroup}\fi
\input amssym
\newfont{\hr}{msbm10}
\newfont{\ams}{msam10}

\font\numbers=cmss12
\font\upright=cmu10 scaled\magstep1
\def\stroke{\vrule height8pt width0.4pt depth-0.1pt}
\def\topfleck{\vrule height8pt width0.5pt depth-5.9pt}
\def\botfleck{\vrule height2pt width0.5pt depth0.1pt}
\def\Zmath{\vcenter{\hbox{\numbers\rlap{\rlap{Z}\kern 0.8pt\topfleck}\kern
2.2pt
                   \rlap Z\kern 6pt\botfleck\kern 1pt}}}
\def\Qmath{\vcenter{\hbox{\upright\rlap{\rlap{Q}\kern
                   3.8pt\stroke}\phantom{Q}}}}
\def\Nmath{\vcenter{\hbox{\upright\rlap{I}\kern 1.7pt N}}}
\def\Cmath{\vcenter{\hbox{\upright\rlap{\rlap{C}\kern
                   3.8pt\stroke}\phantom{C}}}}
\def\Rmath{\vcenter{\hbox{\upright\rlap{I}\kern 1.7pt R}}}
\def\Z{\ifmmode\Zmath\else$\Zmath$\fi}
\def\Q{\ifmmode\Qmath\else$\Qmath$\fi}
\def\N{\ifmmode\Nmath\else$\Nmath$\fi}
\def\C{\ifmmode\Cmath\else$\Cmath$\fi}
\def\R{\ifmmode\Rmath\else$\Rmath$\fi}

\newcounter{app}

\def\app{\setcounter{equation}{0}
\def\theequation{\Alph{app}.\arabic{equation}}\par
   \addvspace{4ex}
   \@afterindentfalse
  \secdef\@app\@dapp}

\newcommand\@app{\@startsection {app}{1}{0ex}%
                                   {-3.5ex \@plus -1ex \@minus -.2ex}%
                                   {2.3ex \@plus.2ex}%
                                   {\normalfont\Large\bf}}
\def\@dapp#1{%
{\parindent \z@ \raggedright  \bf #1}\par\nobreak}
\def\l@app#1#2{\ifnum \c@tocdepth >\z@
    \addpenalty\@secpenalty
    \addvspace{1.0em \@plus\p@}%
    \setlength\@tempdima{8em}%
    \begingroup
      \parindent \z@ \rightskip \@pnumwidth
      \parfillskip -\@pnumwidth
      \leavevmode \bfseries
      \advance\leftskip\@tempdima
      \hskip -\leftskip
      #1\nobreak\hfil \nobreak\hb@xt@\@pnumwidth{\hss #2}\par
    \endgroup\fi}
\newcounter{sapp}[app]

\def\sapp{\def\theequation{\Alph{app}.\arabic{equation}}
\par
\@afterindentfalse
  \secdef\@sapp\@dsapp}
\newcommand{\@sapp}{\@startsection{sapp}{2}{\z@}%
                                     {-3.25ex\@plus -1ex \@minus -.2ex}%
                                     {1.5ex \@plus .2ex}%
                                     {\normalfont\large\bfseries}}

\def\@dsapp#1{%
{\parindent \z@ \raggedright  \bf #1
}\par\nobreak}
\newcommand{\l@sapp}{\@dottedtocline{2}{1.5em}{2.3em}}

\def\stackreb#1#2{\mathrel{\mathop{#2}\limits_{#1}}}
\def\Tr{{\rm Tr}}
\def\res{{\rm res}}
\def\Bf#1{\mbox{\boldmath $#1$}}
\def\balpha{{\Bf\alpha}}

\def\bphi{{\Bf\phi}}
\def\bPhi{{\Bf\Phi}}

\def\d{\partial}
\def\Im{{\rm Im}}

\def\rank{{\rm rank}}
\def\2{{1\over 2}}
\def\N2{${\cal N}=2$}
\def\4N{${\cal N}=4$}

\def\be{ \begin{eqnarray} }
\def\ee{ \end{eqnarray} }
\def\d{\partial}

\def\bea{\begin{eqnarray}}
\def\eea{\end{eqnarray}}
\def\nn{\nonumber}

\def\beq{\begin{equation}}
\def\eeq{\end{equation}}
\def\ba{\beq\new\begin{array}{c}}
\def\ea{\end{array}\eeq}
\def\be{\ba}
\def\ee{\ea}

\begin{document}


\begin{flushright}
FIAN/TD-01/99\\
ITEP/TH-14/99\\
\end{flushright}
\vspace{0.5cm}
\begin{center}
{\LARGE \bf Seiberg-Witten Curves and Integrable Systems
\footnote{Talk given at the Edinburgh conference
"Integrability: the Seiberg-Witten and Whitham Equations", 14-19
September 1998.}}\\
\vspace{1.0cm}
{\Large A.Marshakov
\footnote{e-mail
address: mars@lpi.ac.ru,\ andrei@heron.itep.ru}}\\
\vspace{0.5cm}
{\it Theory
Department, Lebedev Physics Institute, Moscow ~117924, Russia \\
and \\ ITEP, Moscow 117259, Russia}, \\
\end{center}
\bigskip
\begin{quotation}
This talk
gives an introduction into the subject of Seiberg-Witten curves and their
relation to integrable systems. We discuss some motivations and origins of
this relation and consider explicit construction of various families of
Seiberg-Witten curves in terms of corresponding integrable models.
\end{quotation}

\section{Introduction}
\setcounter{equation}{0}

Some time has been already passed after the observation of \cite{GKMMM}
that the effective theories for \N2 vector supermultiplets \cite{SW1,SW2,sun}
can be reformulated in terms of integrable systems. This connection, though
not been clearly understood so far, has become already a beautiful example
of appearance of hidden integrable structure in (multi-dimensional) quantum
gauge theories and, thus, quite a popular topic at many different conferences.
In this talk I was asked to review
basic known facts of this relation and, in addition to brief
explanation of some clear constituents of this correspondence, to present
a list of problems which deserve further investigation.

The formulation of the Seiberg-Witten (SW) solution \cite{SW1,SW2} itself
is very simple: the (Coulomb branch) low-energy effective action for the
4D \N2 SUSY Yang-Mills vector multiplets
(supersymmetry requires the metric on moduli space of massless complex scalars
from \N2 vector supermultiplets to be of "special K\"ahler form" --
or the K\"ahler potential $K({\bf a},{\bar{\bf a}}) = \Im\sum_i {\bar a}_i
{\d{\cal F}\over\d a_i}$ should be expressed through a {\em holomorphic}
function ${\cal F} = {\cal F}(\bf a)$ -- a {\em prepotential})
can be described in terms of auxiliary Riemann surface (complex curve)
$\Sigma $ equipped with meromorphic 1-differential $dS$, possessing peculiar
properties:

\begin{itemize}

\item The number of "live" moduli (of complex structure) of $\Sigma $ is
strongly restricted (roughly "3 times" less than for generic Riemann surface).
The genus of $\Sigma $ for the $SU(N)$ gauge theories is exactly
equal
\footnote{For generic gauge groups one should speak instead of genus -- the
dimension of Jacobian of a spectral curve -- about the dimension of Prym
variety. Practically it means that for other than $A_N$-type gauge theories
one should consider the spectral curves with {\em involution} and only the
invariant under the involution cycles possess physical meaning.
We consider in detail only the $A_N$ theories, the generalization
to the other gauge groups is straightforward: for example, instead of
periodic Toda chains \cite{Toda}, corresponding to $A_N$ theories
\cite{GKMMM}, one has to
consider the "generalized" Toda chains \cite{MW1}, first introduced for
different Lie-algebraic series ($B$, $C$, $D$, $E$, $F$ and $G$) in
\cite{Bogoyav}.}
to the rank of gauge group -- i.e. to the number of independent moduli.

\item The variation of generating 1-form $dS$ over these moduli gives
{\em holomorphic} differentials.

\item The periods of generating 1-form
\be
{\bf a} = \oint_{\bf A}dS
\\
{\bf a}_D = \oint_{\bf B}dS
\label{periods}
\ee
give the set of "dual" masses -- the W-bosons and the monopoles while the
period matrix $T_{ij}(\Sigma)$ -- the set of couplings in the low-energy
effective theory. The prepotential is function of half of the variables
(\ref{periods}), say ${\cal F} = {\cal F}({\bf a})$, then
\be
a_D^i = {\d{\cal F}\over\d a_i}
\\
T_{ij} = {\d a_D^i\over\d a_j} = {\d^2{\cal F}\over\d a_i\d a_j}
\label{adt}
\ee

\end{itemize}

These data mean that the effective SW theory is formulated in
terms of a classical finite-gap integrable system (see, for example,
\cite{DKN} and references therein) and their Whitham deformations
\cite{GKMMM}. The
corresponding integrable models are well-known members of the KP/Toda family
-- for example, pure gauge theory corresponds to a periodic Toda chain
\cite{GKMMM,MW1}, the theories with broken \4N SUSY by the adjoint mass can be
formulated in terms of the elliptic Calogero-Moser models \cite{SWCal},
the theories with extra compact dimension (or Kaluza-Klein modes) give
rise to appearance of relativistic integrable systems
\cite{Ne96,BMMM1,BMMM2}.

The aim of this talk is rather modest -- to give some argumentation in
favour of appearance of complex curves $\Sigma $ in the context of SUSY
gauge theories and to present in clear way how the form of these curves can
be explicitly found in by means of Lax representations of well-known
finite-dimensional integrable systems. I should stress that the curves
$\Sigma $ (and correspondent integrable systems) are {\em auxiliary} from
physical point of view since the quantities (\ref{periods}), (\ref{adt})
effective theories needs depend only on {\em moduli} of SW curve or only on
{\em half} of the variables of classical integrable system. This dependence
is governed by integrable systems which are, in a sense, {\em derivative}
from those we are going to consider below -- the hierarchies of Whitham
and (generalized) associativity equations. Both these subjects are, however,
beyond the scope of these notes.

\section{Motivations and Origins}

In spite of absence at the moment any consistent and full explanations why
the SW curves are identical to the curves of integrable systems, let us start
from some physical motivations. We consider, first, the perturbative limit
of \N2 SUSY gauge theories and show that (degenerate) spectral curves and
corresponding integrable systems appear already at this level. The situation
is much more complicated for the non-perturbative picture and the only
way to explain the origin of full SW curve exists in the framework of
non-perturbative string theory or M-theory. In the second part of this
section we shall discuss briefly how the Lax representations of the SW
spectral curves arise in this context.

\subsection{Perturbative spectral curves}

Amazingly enough the relation between SW theories and integrable systems can
be discussed already at the {\em perturbative} level, where \N2 SUSY
effective actions are completely defined by the 1-loop contributions (see
\cite{SW1} and references therein). The scalar field $\bPhi = \|\Phi_{ij}\|$
of \N2 vector supermultiplet acquires nonzero VEV $\bPhi = {\rm
diag}(\phi_1,\dots,\phi_N)$ and the masses of ``particles'' --
$W$-bosons and their superpartners are proportional to
$\phi_{ij}=\phi_i-\phi_j$ due to
the Higgs term $[A_\mu,\bPhi]_{ij} = A_\mu^{ij}(\phi_i-\phi_j)$ in the SUSY
Yang-Mills action. These masses can be written altogether in terms of the
generating polynomial
\be
w = P_N(\lambda ) = \det (\lambda - \bPhi) = \prod (\lambda - \phi_i)
\label{polyn}
\ee
($\bPhi $ -- the adjoint complex scalar, $\Tr\bPhi = 0$,)
via residue formula
\be\label{pertsample}
m_{ij} \sim \oint_{C_{ij}} \lambda d\log w=
\oint_{C_{ij}} \lambda d\log P_{N}(\lambda )
\ee
which for a particular "$\infty$-like" contour $C_{ij}$ around the roots
$\lambda = \phi_i$ and $\lambda = \phi_j$ gives rise to the Higgs masses. The
contour integral (\ref{pertsample}) is defined on a complex plane --
$\lambda $-plane with $N$ removed points: the roots of the polynomial
(\ref{polyn}) -- a {\em degenerate} Riemann surface. The masses of monopoles
are naively infinite in this limit, since the corresponding contours (dual
to $C_{ij}$) start and end in the points where $dS$ obeys pole
singularities. It means that the monopole masses, proportional to the
squared inverse coupling, are renormalized in perturbation theory and defined
naively up to the masses of particle states times some divergent constants.

The effective action (the prepotential) ${\cal F}$, or the set of effective
charges $T_{ij}$ (\ref{adt}), are defined in \N2 perturbation theory by
1-loop diagram giving rise to the logarithmic contribution
\be\label{effcharge}
\left(\delta ^2{\cal F}\right)_{ij} = T_{ij}
 \sim \sum_{\rm masses} \log {({\rm mass})^2\over\Lambda ^2} =
\log {(\phi_i-\phi_j)^2\over\Lambda ^2}
\ee
where $\Lambda\equiv\Lambda_{QCD}$ and last equality is written only for pure
gauge theories -- since the only masses we have there are the Higgs masses
(\ref{pertsample}). That is all one has
in the perturbative {\em weak-coupling} limit of the SW construction, when
the instanton contributions to the prepotential (being proportional to
the degrees of  $\Lambda^{2N}$ (or $q^{2N}\equiv e^{2\pi i\tau N}$ -- in the
UV-finite theories with bare coupling $\tau $)
are (exponentially) suppressed so that one keeps only the terms proportional
to $\tau $ or $\log\Lambda$. We shall list several more examples below and
demonstrate that these degenerated {\em rational} spectral curves can
be already related to the family of {\em trigonometric}
Ruijsenaars-Schneider and Calogero-Moser-Sutherland systems and the {\em
open} Toda chain or Toda molecule.

Start with the original case of $SU(2)$ pure gauge theory. Eq.~(\ref{polyn})
turns into
\be
w = \lambda^2 - u
\label{polsu2}
\ee
with $u = \2\Tr\bPhi^2$. In the parameterization of \cite{SW1}
$X = w = e^z = \lambda^2 - u$, $Y = w\lambda$ the same equation can be
written as
\be
Y^2 = X^2(X+u)
\label{optoda}
\ee
and the masses (\ref{pertsample}) are now defined by the contour integrals of
\be
dS = \lambda d\log w = 2{\lambda^2d\lambda\over\lambda^2-u}
= {\lambda d\lambda\over\lambda - \sqrt{u}} +
{\lambda d\lambda\over\lambda + \sqrt{u}}  = \sqrt{X+u}~{dX\over X}
\label{dSop}
\ee
One can easily notice that Eqs.~(\ref{polsu2}), (\ref{optoda}) and
(\ref{dSop}) can be interpreted as {\em integration} of the open
$SL(2)$ (the Liouville) Toda chain with the co-ordinate $X=w=e^q$, momentum
$p=\lambda$ and Hamiltonian (energy) $u$. The integration of generating
differential $dS = pdq$ over the trajectories of the particles gives rise,
in fact, to the monopole masses in the SW theory.

This is actually a general rule -- the perturbative \N2 theories of the
"SW family" give rise to the "open" or trigonometric family of integrable
systems -- the open Toda chain, the trigonometric Calogero-Moser or
Ruijsenaars-Schneider systems. This can be easily established
at the level of spectrum (\ref{pertsample}) and the effective couplings
(\ref{effcharge}) -- the corresponding (rational) curves are (\ref{polyn})
in the $N$-particle Toda chain case
\be
w = {P^{(CM)}_N(\lambda )\over P^{(CM)}_N(\lambda + m)}
\ \ \ \ \ \ \ \ \
dS = \lambda{dw\over w}
\label{triCa}
\ee
for the trigonometric Calogero-Moser-Sutherland model and
\be
w=\frac{P^{(RS)}_N(\lambda)}
{P^{(RS)}_N(\lambda e^{-2i\epsilon})}
\ \ \ \ \ \ \ \ \
dS = \log\lambda {dw\over w}
\label{triRu}
\ee
for the trigonometric Ruijsenaars-Schneider system. It is easy to see that
(perturbative) spectra are given by general formula \cite{BMMM2}
\be
M = \phi_{ij} \oplus {\pi n\over R} \oplus {\epsilon +\pi n\over R}
\\
N\in {\bf Z}.
\label{spectrum}
\ee
and contain in addition to the Higgs part $\phi_{ij}$ the Kaluza-Klein (KK)
modes $\pi n\over R$ and the KK modes for the fields with "shifted" by
$\epsilon$ boundary conditions. The $\epsilon$ parameter can be treated as a
Wilson loop of gauge field along the compact dimension -- a kind of
different (or dual) moduli in the theory and in a subclass of models it
plays the role of the mass of the adjoint matter multiplet.

\subsection{Nonperturbative SW Curves, M-Theory and $\bar{\partial}$-equation}

As an example, let us consider, first, the case of pure $SU(2)$ \N2 gauge
theory \cite{SW1}.
The gauge group has rank 1 and from the "integrable" point of view the
situation is trivial, since the correspondent
integrable model is "one-dimensional" (phase space has dimension two) and
this case can be always solved explicitly.
The full ("blown-up") spectral curve has
the form
\be\label{su2}
w + {\Lambda^4\over w}\ \stackreb{w=\Lambda^2e^z}{=}\
2{\Lambda ^2}\cosh z = \lambda ^2 - u \equiv P_2(\lambda)
\ee
(cf. with (\ref{polyn}))
and coincides with the equation relating the Hamiltonian (energy) $u$
with the co-ordinate $z=i q$ (or $z= q$) and momentum $\lambda = p$ of a
particle moving into the $SL(2)$ Toda-chain potential --
the well-known physical pendulum, instead of considered previously
"Liouville wall"
\footnote{Note, that for the SW theory one has to consider {\em both}
phases -- the sine-Gordon and the sinh-Gordon -- of an integrable system
together, that is why sometimes people speak in this respect about a
{\em complex} integrable system.}.
There are two other (hyper)elliptic parameterizations of the
$SU(2)$ SW curve: the first one was proposed in original paper \cite{SW1}
\be
Y^2 = (X^2 -4\Lambda^4)(X+u)
\\
X = P_2(\lambda ) = \lambda^2 - u \ \ \ \ \ \ \ Y = \lambda y
\label{sw1}
\ee
with $dS = (X+h){dX\over Y}$, $dt = {dX\over Y}$ and another one used
in \cite{SW3}
\be
{\tilde y}^2 = w^3 + hw^2 + \Lambda^4 w
\\
w = \Lambda^2e^z \ \ \ \ \ \ \ {\tilde y} = \lambda w
\label{sw3}
\ee
endowed with $dS = {\tilde y}{dw\over w} = {(w^2+uw+1)dw\over w{\tilde y}}$
and $dt = {dw\over\tilde y}$.
Since {\em any} 1-dimensional system with conserving energy is integrable,
(\ref{su2}) gives
\be\label{holsu2}
dt = {d q\over p} = {dz\over\lambda} = 2{d\lambda\over y}
\ee
where
\be
y^2 = (\lambda^2 - u)^2 - 4\Lambda^4 = 4\sinh^2z
\label{hyelsu2}
\ee
so that
\be
t = \int {dX\over Y} = 2\int {d\lambda\over y}
\ee
is the Abel map for (\ref{sw1}) or (\ref{su2}).
The (normalized) action integrals in two different ("sin- and sinh- Gordon")
phases are the periods
\be\label{asu2}
(a,a_D) = \oint_{(A,B)} dS = \oint_{(A,B)} pd q =
\oint_{(A,B)}  d q\sqrt{u + \Lambda^2\cos  q}
\ee
In the "$\sin$-phase" if $u >> \Lambda^2$
the interaction is inessential and, the main effect for the integral
(\ref{asu2}) comes from
\be
a \sim \sqrt{u}\int_A d q \sim \sqrt{u}\times const
\label{abig}
\ee
In another ``phase'', corresponding to {\em imaginary} values
of $ q$ ($ q \rightarrow i q + \pi$) the action integral
(\ref{asu2}) turns into
\be\label{adsu2}
a^D = \oint_B pd q = \oint_B d q\sqrt{u - \Lambda^2\cosh  q}
\ee
and in the first approximation is equal to
\be
a^D \sim \int_B d q\sqrt{u-\Lambda^2\cosh  q} \sim \sqrt{u}\int_B d q
\sim \sqrt{u}\times  q_* = \sqrt{u}\log{u\over\Lambda^2}
\label{adbig}
\ee
where we by $ q_*$ we have denoted the ``turning point''
$\Lambda^2e^{ q_*}=u$.
It is easy to see that the expressions (\ref{abig}) and (\ref{adbig})
give the perturbative values of the W-boson $m_W^2=u$ and monopole $m_M^2=
\left({m_W\over g_{YM}^2}\right)^2$ masses in
the $SU(2)$ SW theory.  The prepotential for $u >> \Lambda^2$ is thus
\be
{\cal F} = {u\over 2}\left(\log{u\over\Lambda^2} +
o\left( {\Lambda^2\over u}\right)\right)
\label{prepbig}
\ee
This is an obvious perturbative (1-loop) result of \N2 SUSY Yang-Mills
theory. The main hypothesis of \cite{SW1} is that formulas (\ref{asu2}) (and
the prepotential or the set of coupling constants they define) are valid
{\em beyond} the perturbation theory. In this case the r.h.s. of
(\ref{prepbig}) would contain an infinite instanton series (in
${\Lambda^4\over u^2}$ for the $SU(2)$ gauge group) but this can be encoded
into a relatively simple modification of SW curves.

In the construction inspired by M-theory moduli arise either as positions of
the D-branes $\phi_i \sim {|r_i|\over\alpha '}$ or as monodromies  of the
gauge fields $\epsilon = \oint A_Mdx^M$ along the compactified
directions (and are related to the positions of branes by T-duality), i.e.
they are given by the set of data $(A_M,\bPhi^{(i)})$.
The perturbative spectral curves (\ref{polsu2}), (\ref{triCa}), (\ref{triRu})
correspond in this picture to the $p$-branes ($p$-dimensional hypersurfaces
with the $(p+1)$-dimensional world volume) with the D$(p-1)$-brane sources
\cite{WitM}. In particular case when the
configuration can be described by {\em two} real (or one complex
component) of both kind of fields $(\bar{A},\bPhi)$ one can define the full
(smooth) spectral curves $\Sigma_g$ as a cover of some bare spectral curve
$\Sigma_0$ -- usually a torus, which can be naturally chosen when one has at
least two compactified dimensions --
by generalization of the equation (\ref{polyn})
\be
\det(\lambda-\bPhi (z)) = 0
\label{laxeq}
\ee
where $\bPhi =\bPhi (z) $ is now function (in fact 1-differential) on bare
curve $\Sigma_0$ and obeys \cite{Hi,GoNe}
\be
{\bar\partial}\Phi + [{\bar A},\Phi ]=\sum_{\alpha}J^{(\alpha )}
\delta^{(2)}(P-P_{\alpha})
\label{gauss}
\ee
i.e. is {\em holomorphic} in the complex structure determined by $\bar{A}$.
The invariants of $\bar A$ can be thought of as co-ordinates
(one commuting set of variables) while the invariants of $\Phi$ as
hamiltonians (another commuting set of variables) of an integrable system.
It means that M-theory point of view implies that the VEV $\bPhi $ becomes a
function on some base spectral curve $\Sigma_0$ -- usually a cylinder or
torus, appearing as a part of the (compactified) brane world volume and
satisfies $\bar{\partial}$-equation. Such holomorphic (or, better,
meromorphic) objects were introduced long ago \cite{DKN} as {\em Lax
operators} for the finite-dimensional integrable systems, holomorphically
depending on some spectral parameter.

The (first-order) equation (\ref{gauss}) arises from the BPS-like
condition \cite{Diac,MMaM} of the type $Q\psi = \Gamma_MD_M\Phi +
\Gamma_{MN}F_{MN} = 0$ which determines the form of the Lax operator $\Phi$
and, thus, the shape of the curve (\ref{laxeq}). On torus with $p$ marked
points $z_1,\dots ,z_p$ can be defined by ($i,j=1,\dots,N$)
\be\label{hi-ell}
\bar\partial\Phi _{ij} + (q_i-q_j)\Phi _{ij} =
\sum _{\alpha =1}^p J_{ij}^{(\alpha )}\delta (z - z_{\alpha})
\ee
so that the solution has the form ($q_{ij}\equiv q_i-q_j$)
\footnote{As usual, by $\theta_{\ast}(z|\tau)\equiv\theta_{11}(z|\tau)$ the
(only on torus) {\em odd} ($\theta_{\ast}(0|\tau)=0$) theta-function
is denoted.}
\be
\Phi _{ij}(z) = \delta _{ij}\left(p_i +
\sum _{\alpha}J_{ii}^{(\alpha )}\partial\log\theta (z-z_{\alpha}|\tau )\right)
+
\\
+ \left( 1-\delta _{ij}\right) e^{q_{ij}(z-{\bar z})}
\sum _{\alpha}J_{ij}^{(\alpha )}
{\theta (z - z_{\alpha} + {\Im\tau\over\pi}q_{ij})\theta '(0)\over
\theta (z - z_{\alpha})\theta ({\Im\tau\over\pi}q_{ij})}
\ee
The exponential (nonholomorphic) part can be removed by
a gauge transformation
\be\label{laxgauge}
\Phi _{ij}(z)\rightarrow (U^{-1}\Phi U)_{ij}(z)
\ee
with $U_{ij} = e^{q_i{\bar z}}\delta_{ij}$.
The additional conditions to the matrices $J_{ij}^{(\alpha )}$
\be
\sum _{\alpha = 1}^p J^{(\alpha )}_{ii}=0
\ee
imply that the sum of all residues of a function
$\Phi _{ii}$ vanishes, and
\be\label{masses-par}
\Tr J^{(\alpha )} = m_{\alpha}
\ee
with $m_{\alpha} = {\rm const}$ being some parameters ("masses") of a theory.
The spectral curve equation becomes
\be\label{gencu}
{\cal P}(\lambda ;z) \equiv
\det _{N\times N}\left(\lambda - \bPhi (z)\right) = \lambda ^N
+ \sum _{k=1}^N\lambda ^{N-k}f_k(z) = 0
\ee
where $f_k(z)\equiv f_k(x,y)$ are some functions (in general with $k$ poles)
on the elliptic curve. If, however, $J^{(\alpha )}$ are restricted
by
\be
\rank J^{(\alpha )} \leq l\ \ \ \ \ \ l<N
\ee
the functions $f_k(z)$ will have poles at $z_1,\dots,z_p$ of the order not
bigger than $l$.
The generating differential, as usual, should be
\be\label{dS1}
dS = \lambda dz
\ee
and its residues in the marked points
$(z_{\alpha}, \lambda ^{(i)}(z_{\alpha}))$ (different $i$ correspond to the
choice of different sheets of the covering surface) are related with the
mass parameters (\ref{masses-par}) by
\be
m_{\alpha} = \res _{z_{\alpha}}\lambda dz \equiv
\sum _{i=1}^N \res _{\pi _{(i)}^{-1}z_{\alpha}}\lambda ^{(i)}(z)dz
= \res _{z_{\alpha}}\Tr\bPhi dz
\ee
We shall see that general form of the curve (\ref{gencu}) coincides
with the general curves arising in the SW theory (in many important cases
torus should degenerate into a cylinder). For example, the $p=1$, $l=1$
case gives rise to the elliptic Calogero-Moser model
(see eq.~(\ref{LaxCal}) below).

Thus, in the M-theory picture it becomes clear that full
non-perturbative SW curves are smooth analogues of their degenerate
perturbative cousins. This blowing up corresponds to the "massive"
deformation of previous family of integrable systems, or, more strictly, the
non-perturbative SW curves correspond to the family of periodic Toda chains
\cite{Toda} and Calogero-Moser-Ruijsenaars \cite{Cal,KriCal,RS,R} integrable
models.

\section{Zoo of curves and integrable systems}

Now let us turn to the question of the zoo of SW's
integrable systems -- or some classification of relations between the SUSY
YM theories and corresponding integrable models. We shall consider the cases
of broken \4N SUSY theories, or \N2 SUSY Yang-Mills with extra {\em adjoint}
matter multiplets, theories with soft Kaluza-Klein (KK) modes or with extra
5th compact dimension and, as a separate question, theories with fundamental
matter (all  with the $SU(N)$ gauge group).
The last issue is relatively less investigated yet, i.e. there still
exist some open problems, mostly related with the "conformal" case $N_f=2N$.
Two first classes instead can be formulated in a unique way since there exists
a "unifying" integrable system -- the elliptic
Ruijsenaars-Schneider model \cite{RS,R} (with bare coupling $\tau$, "relativistic"
parameter $R$ -- the radius of compact dimension and an extra parameter
$\epsilon $ -- see (\ref{spectrum})), giving rise to all known models of
these two classes in its various degenerations \cite{BMMM2}:

\begin{itemize}
\item If $R\rightarrow 0$ (with finite $\epsilon$) the
mass spectrum (\ref{spectrum}) reduces to a single point
$M = 0$, i.e. all masses arise only due to the Higgs effect. This is the
standard
four dimensional \N2 SUSY YM model associated with the periodic Toda chain.
In this situation \N2 SUSY in four dimensions is insufficient to ensure
UV-finiteness, thus bare coupling diverges $\tau \rightarrow i\infty$, but
the dimensional transmutation substitutes the dimensionless
$\tau$ by the new dimensionful (and {\em finite}) parameter $\Lambda^N =
e^{2\pi i\tau}  (\epsilon/R)^{N}$.

\item If $R \rightarrow 0$ and $\epsilon \sim mR$ for finite $m$,
then UV finiteness is preserved. The mass spectrum
reduces to the two points $M = 0$ and $M = m$. This is the
four dimensional YM model with \4N SUSY softly broken to \N2.
The associated finite-dimensional integrable system \cite{SWCal} is the elliptic
Calogero-Moser model \cite{Cal,KriCal}. The previous case is then obtained by
Inosemtsev's \cite{Ino} double scaling limit
when $m\rightarrow 0$, $\tau \rightarrow i\infty$ and
$\Lambda^N = m^N e^{2\pi i\tau}$ is fixed.

\item If $R \neq 0$ but $\epsilon \rightarrow i\infty$ the
mass spectrum reduces to a single
Kaluza-Klein tower, $M = \pi n/R$, $n\in Z$. This compactification of
the five dimensional model has $N=1$ SUSY and is not
UV-finite. Here $\tau \rightarrow i\infty$ and
$\epsilon \rightarrow i\infty$, such that $2\pi\tau - N\epsilon$
remains finite. The corresponding integrable system \cite{Ne96} is the
relativistic Toda chain \cite{relToda}.

\item Finally, when $R\neq 0$ and $\epsilon$ and $\tau$
are both finite one distinguished case still remains:
$\epsilon =\pi/2$.\footnote{The case $\epsilon = 0$
of fully unbroken five dimensional $N=2$ supersymmetry
is of course also distinguished, but trivial:
there is no evolution of effective couplings (renormalization
group flows) and the integrable system is just that of $N$
non-interacting (free) particles.}
Here only periodic and anti periodic boundary conditions
occur in the compact dimension. This is the case analyzed in
\cite{BMMM1,BMMM2} and interested reader can find all details there.

\end{itemize}

\subsection{Toda chain}

This is the most well-known example of $SU(N)$ pure gauge theory \cite{sun}.
The spectral curve is
\be\label{sun}
w + {\Lambda^{2N}\over w} = P_N(\lambda )
\\
P_N(\lambda ) \equiv \det (\lambda - \bPhi ) = \lambda^N - \sum u_k\lambda^k
\ee
and the generating 1-form $dS = \lambda{dw\over w}$ satisfies
\be
\delta_{\rm moduli}dS \equiv \left.\delta_{\rm moduli}dS \right|_{w=const}
= \left(\delta_{\rm moduli}\lambda\right){dw\over w} =
{\sum\lambda^k \delta u_k \over P_N'(\lambda)}{dw\over w} =
\sum{\lambda^kd\lambda\over y}\delta u_k = {\rm holomorphic}
\label{hol}
\ee
where
\be
y^2 = \left(w - {\Lambda^{2N}\over w}\right)^2
= P_N^2(\lambda ) - 4\Lambda^{2N}
\\
P_N'(\lambda ) \equiv\left. {\d P_N\over\d\lambda }\right|_{u=const}
\ee
The spectral curve (\ref{sun}) corresponds to periodic $N$-particle problem in
Toda chain \cite{DateTa,KriDu}. Let us demonstrate now how its form can be
derived in the language of integrable systems
using two different forms of the Lax representation with spectral
parameter for periodic Toda chain.

The Toda chain system \cite{Toda,UT84} is a system of
particles with only the neighbors pairwise exponential interaction
with the equations of motion
\footnote{For simplicity in this section we consider a periodic
Toda chain with coupling constant equal to unity. It is easy to restore it
in all the equations; then it becomes clear that it should be identified with
$\Lambda = \Lambda _{QCD}$ -- the scale parameter of pure \N2 SUSY
Yang-Mills theory.}
\be\label{Todaeq}
\frac{\partial  q_i}{\partial t} = p_i \ \ \ \ \
\frac{\partial p_i}{\partial t} = e^{ q_{i+1} - q_i}-
e^{ q_i- q_{i-1}}
\ee
where one assumes (for the periodic problem with the ``period" $N$) that
$ q_{i+N} =  q_i$ and
$p_{i+N} = p_i$. It is an integrable system, with $N$
Poisson-commuting Hamiltonians, $h_1 = \sum p_i$ = P, $h_2 =
\sum\left(\frac{1}{2}p_i^2 + e^{ q_i- q_{i-1}}\right) = E$, etc.
Starting from naively
infinite-dimensional system of particles (\ref{Todaeq}), the periodic
problem can be formulated
in terms of (the eigenvalues and the eigenfunctions of) two commuting
operators: the Lax operator ${\cal L}$ (or the auxiliary linear problem for
(\ref{Todaeq})
\footnote{Equation (\ref{laxtoda}) is the second-order difference equation
and it has two independent solutions which we shall often denote below as
$\Psi^+$ and $\Psi^-$. In more general framework of the Toda lattice hierarchy
these two solutions correspond to the two possible choices of sign in the
time-dependent form of the Lax equation (\ref{laxtoda})
\be
\pm {\d\over\d t}\Psi^{\pm}_n = \sum_k {\cal L}_{nk}\Psi^{\pm}_k
\label{laxtoda1}
\ee})
\be\label{laxtoda}
\lambda\Psi _n =
\sum _k {\cal L}_{nk}\Psi _k =
e^{{1\over 2}( q_{n+1}- q_n)}\Psi _{n+1} + p_n\Psi _n +
e^{{1\over 2}( q_n- q_{n-1})}\Psi _{n-1}
\ee
and the second operator -- in our case of
periodic boundary conditions to be chosen as a {\it monodromy} or shift
operator in a discrete variable -- the number of a particle
\be\label{T-op}
T q_n =  q_{n+N}\ \ \ \ \ \ Tp_n = p_{n+N}\ \ \ \ \ \ \ \
T\Psi_n = \Psi_{n+N}
\ee
The existence of common spectrum of these two operators
\footnote{Let us point out that we consider a {\it periodic} problem for
the Toda chain when only the BA function can acquire a nontrivial factor
under the action of the shift operator while the coordinates and momenta
themselves are periodic. The {\it quasi}periodicity of coordinates and momenta
-- when they acquire a nonzero shift -- corresponds to the change of
the coupling constant in the Toda chain Hamiltonians.}
\be\label{spec}
{\cal L}\Psi = \lambda\Psi \ \ \ \ \ T\Psi = w\Psi\ \ \ \ \ \ [{\cal L},T]=0
\ee
means that there is a relation between them ${\cal P}({\cal L},T) = 0$
which can be strictly formulated in terms of spectral curve
\cite{Krico} $\Sigma$:
${\cal P}(\lambda,w) = 0$. Here is an important difference with the
perturbative case considered above, where only one (Lax) operator was
really defined and there was no analog of the second $T$-operator.
The generation function for the integrals of motion -- the Toda chain
Hamiltonians can be written in terms of
${\cal L}$ and $T$ operators and Toda chains possess two
different (though, of course, equivalent) formulations of this kind.

In the first version the Lax operator (\ref{laxtoda}) should be
re-written in the
basis of the $T$-operator eigenfunctions and becomes the $N\times N$
matrix,
\be\label{LaxTC}
{\cal L} = {\cal L}(w) = {\bf pH} +
\sum _{{\rm simple}\ \alpha}e^{\balpha\bphi}\left( E_{\alpha}
+ E_{-\alpha}\right) + w^{-1}e^{-\balpha _0\bphi}E_{-\alpha _0} +
w e^{-\balpha _0\bphi}E_{\alpha _0} =
\\ =
\left(\begin{array}{ccccc}
 p_1 & e^{{1\over 2}( q_2- q_1)} & 0 & & we^{{1\over 2}
( q_1- q_{N})}
\\
e^{{1\over 2}( q_2- q_1)} & p_2 & e^{{1\over 2}( q_3 -  q_2)} &
\ldots & 0\\
0 & e^{{1\over 2}( q_3- q_2)} & p_3 & & 0 \\
 & & \ldots & & \\
\frac{1}{w}e^{{1\over 2}( q_1- q_{N})} & 0 & 0 & & p_{N}
\end{array} \right)
\ee
defined on a cylinder, i.e. it explicitly depends on spectral
parameter $w$ -- the eigenvalue of the shift operator (\ref{T-op}).
Matrix (\ref{LaxTC}) is almost exactly
three-\-diagonal, the only extra nonzero
elements appear in the off-diagonal corners due to the periodic
conditions (\ref{T-op}) reducing in this way naively infinite-dimensional
constant matrix (\ref{laxtoda}) to a finite-dimensional one, but depending on
spectral parameter $w$.
The eigenvalues of the Lax operator (\ref{LaxTC}) are defined from the spectral
equation
\be\label{SpeC}
{\cal P}(\lambda,w) = \det_{N\times N}\left({\cal L}^{TC}(w) -
\lambda\right) = 0
\ee
Substituting the explicit expression (\ref{LaxTC}) into (\ref{SpeC}),
one gets:
\be\label{fsc-Toda}
{\cal P}(\lambda,w) = w + \frac{1}{w} - P_{N}(\lambda ) = 0
\ee
i.e. eq.~(\ref{sun}), where $P_{N}(\lambda )$ is a polynomial of degree $N$,
with the mutually Poisson-commuting coefficients:
\be\label{Schura}
P_{N}(\lambda ) = \lambda ^{N} + h_1 \lambda ^{N-1} +
\frac{1}{2}(h_2-h_1^2)\lambda ^{N-2} + \ldots
\ee
($h_k = \sum_{i=1}^{N} p_i^k + \ldots$),
parameterizing (a subspace in the) moduli space of the complex structures of
the hyperelliptic curves $\Sigma^{TC}$ of genus $N - 1 = \rank\ SU(N)$.

An alternative description of the same system arises when one ({\em before}
imposing the periodic conditions!) {\it solves}
explicitly the auxiliary linear problem (\ref{laxtoda}) which is a
second-order difference equation. To solve it one rewrites
(\ref{laxtoda}) as
\be\label{recrel}
\Psi _{i+1} = (\lambda - p_i)\Psi _i - e^{ q_i - { q_{i+1} +  q_{i-1}
\over 2}}
\Psi_{i-1}
\ee
or, since the space of solutions is 2-dimensional
\footnote{The initial condition for the recursion relation (\ref{recrel})
consists of two arbitrary functions, say, $\Psi _1$ and $\Psi _2$, which
are, of course, linear combinations of $\Psi ^{+}$
and $\Psi ^{-}$ from (\ref{laxtoda1}).},
it can be rewritten as ${\widetilde\Psi}_{i+1}
= L_i(\lambda ){\widetilde\Psi}_i$ where $\widetilde\Psi _i$ is a set of
two-vectors and $L_i$ -- a chain of $2\times 2$ Lax matrices.
After a simple "gauge" transformation $L_i \rightarrow
U_{i+1}L_iU_i^{-1}$ where $U_i = {\rm diag}(e^{\2 q_i},e^{-\2 q_{i-1}})$
(and replacing $p_i \rightarrow - p_i$) these matrices can be written in the
form \cite{FT}
\be\label{LTC}
L_i(\lambda) =
\left(\begin{array}{cc} \lambda + p_i & e^{ q_i} \\ e^{- q_i} & 0
\end{array}\right), \ \ \ \ \ i = 1,\dots ,N
\ee
The matrices (\ref{LTC}) obey {\it quadratic}
$r$-matrix Poisson bracket relations: the canonical Poisson
brackets of the co-ordinates and momenta of the Toda chain particles
$\{  q_i,p_j\}= \delta _{ij}$ can be equivalently rewritten in the
"commutator" form
\be\label{quadrP} \left\{
L_i(\lambda)\stackrel{\otimes}{,}L_j(\lambda')\right\} =
\delta_{ij} \left[ r(\lambda - \lambda'), L_i(\lambda)\otimes
L_j(\lambda')\right]
\ee
with the ($i$-independent!) numerical rational
$r$-matrix $r(\lambda) = \frac{1}{\lambda } \sum_{a=1}^3 \sigma_a\otimes
\sigma^a$ satisfying the classical Yang-Baxter equation.
As a consequence, the transfer matrix
\be\label{monomat}
T_{N}(\lambda) =
\prod_{N \ge i\ge 1}^{\curvearrowleft} L_i(\lambda )
\ee
satisfies the same Poisson-\-bracket relation
\be
\left\{ T_{N}(\lambda)\stackrel{\otimes}{,}T_{N}(\lambda')\right\}
= \left[ r(\lambda - \lambda'), T_{N}(\lambda)\otimes
T_{N}(\lambda')\right]
\label{quadrT}
\ee
and the integrals of motion of the Toda chain are generated
by another representation of spectral equation
\be\label{specTC0}
\det_{2\times 2}\left( T_{N}(\lambda )
- w\right) = w^2 - w\Tr T_{N}(\lambda ) + \det T_{N}(\lambda ) =
\\
= w^2 - w\Tr T_{N}(\lambda ) + 1 = 0
\ee
(using that $\det_{2\times 2} L(\lambda) = 1$ leads to $\det_{2\times 2} T_{N}
(\lambda) = 1$), or
\be\label{specTC}
{\cal P}(\lambda ,w) =  w + \frac{1}{w} - \Tr T_{N}(\lambda)=
w + \frac{1}{w} - P_{N}(\lambda)= 0
\ee
which coincides with (\ref{fsc-Toda}).
The polynomial $P_{N}(\lambda)=\Tr T_{N}(\lambda)$ in (\ref{specTC})
is of degree $N$, its coefficients are the integrals of motion
since
\be\label{trcom}
\left\{ \Tr T_{N}(\lambda ), \Tr T_{N}(\lambda' )\right\} =
\Tr \left\{ T_{N}(\lambda )\stackrel{\otimes}{,}T_{N}(\lambda' )\right\} =
\nn \\
= \Tr \left[ r(\lambda - \lambda' ), T_{N}(\lambda )\otimes
T_{N}(\lambda' )\right] = 0
\ee
Finally, let us note that the Toda chain $N\times N$ Lax
operator (\ref{LaxTC}) by gauge transformation can be brought to another
familiar form
\be\label{LaxTCHi}
{\cal L}^{TC}(w) \rightarrow \tilde{\cal L}^{TC} = U^{-1}{\cal L}^{TC}(w)U=
\nn \\
= {\bf pH} + v^{-1}\left( e^{-\balpha _0\bphi}E_{-\alpha _0} +
\sum _{{\rm simple}\ \alpha}e^{\balpha\bphi}E_{\alpha}\right)+
 v\left( e^{-\balpha _0\bphi}E_{\alpha _0} +
\sum _{{\rm simple}\ \alpha}e^{\balpha\bphi}E_{-\alpha}\right) =
\\
= \left(\begin{array}{ccccc}
 p_1 & \frac{1}{v}e^{{1\over 2}( q_2- q_1)} & 0 &
 & ve^{{1\over 2}( q_1- q_{N})}\\
ve^{{1\over 2}( q_2- q_1)} & p_2
& \frac{1}{v}e^{{1\over 2}( q_3 -  q_2)} & \ldots & 0\\
0 & ve^{{1\over 2}( q_3- q_2)} & p_3 & & 0 \\
 & & \ldots & & \\
\frac{1}{v}e^{{1\over 2}( q_1- q_{N})} & 0 & 0 & & p_{N}
\end{array} \right)
\nn \\
U_{ij} = v^i\delta _{ij} \ \ \ \ \ w\equiv v^{N}
\ee
Formally this corresponds to change of gradation of the Toda chain Lax
operator, the form (\ref{LaxTCHi}) is especially natural relates the
Toda chain Lax operator with the $\widehat{sl(N)}$ Kac-Moody algebra.
In the form (\ref{LaxTCHi}) the periodic Toda chain can be thought of
as a special
"double-scaling" limit of the $SL(N)$ Hitchin system on torus with a marked
point -- the Calogero-Moser model (see below). It is clear that the Lax
operator (\ref{LaxTCHi}) satisfies the $\bar{\partial}$-equation
(\ref{gauss}) on a cylinder with {\em trivial} gauge connection \cite{M97}
\be\label{hito}
\bar\partial _v \tilde{\cal L}^{TC}(v) =
\left( e^{-\balpha _0\bphi}E_{-\alpha _0} +
\sum _{{\rm simple}\ \alpha}e^{\balpha\bphi}E_{\alpha}\right)
\delta (P_0) - \\ -
\left( e^{-\balpha _0\bphi}E_{\alpha _0} +
\sum _{{\rm simple}\ \alpha}e^{\balpha\bphi}E_{-\alpha}\right)
\delta (P_{\infty})
\ee
and they can be easily solved giving rise to (\ref{LaxTCHi}).

\subsection{Broken N=4 SUSY and the Elliptic Calogero-Moser Model
\label{ss:matter}}

Now, let us turn to the observation that
the $N\times N$ matrix Lax operator (\ref{LaxTC}) can be thought of as a
"degenerate" case of the Lax operator for the $N$-particle Calogero-Moser
system \cite{KriCal}
\be\label{LaxCal}
{\cal L}^{CM}(z) =
\left({\bf pH} + \sum_{\balpha}F({\bf q\balpha}|z)
E_{\balpha}\right) = \nn \\
= \left(\begin{array}{cccc}
 p_1 & F(q_1-q_2|z) & \ldots &
F(q_1 - q_{N}|z)\\
F(q_2-q_1|z) & p_2 & \ldots &
F(q_2-q_{N}|z)\\
 & & \ldots  & \\
F(q_{N}-q_1|z) &
F(q_{N}-q_2|z)& \ldots &p_{N}
\end{array} \right)
\ee
The matrix elements $F(q|z) = m\frac{\sigma(q+z)}{\sigma(q)\sigma(z)}
e^{\zeta(q)z}$ are expressed in terms of the Weierstrass sigma-functions
so that the Lax operator ${\cal L}(z)$ is defined on elliptic curve
$E(\tau)$
\be\label{ell1}
y^2 = (x -  e_1)(x - e_2)(x - e_3)
\\
x = \wp(z)\ \ \ \ \ \ \ \ \ y = \2\wp'(z)
\ee
or a complex torus with modulus $\tau $ and one marked point
$z=0, x=\infty, y=\infty$. The Lax operator (\ref{LaxCal}) corresponds to
completely
integrable system with Hamiltonians $h_1 = \sum_i p_i = P$, $h_2 =
\sum_i p_i^2 + m^2\sum_{i<j}\wp (q_i - q_j)$, etc; the coupling constant
$m$ in 4D interpretation plays the role of the mass of adjoint matter
${\cal N}=2$ hypermultiplet breaking ${\cal N}=4$ SUSY down to ${\cal N}=2$
\cite{SWCal}.

>From Lax representation (\ref{LaxCal}) it follows that the spectral curve
$\Sigma^{CM}$ for the $N$-particle Calogero-Moser system
\be\label{fscCal}
\det_{N\times N} \left({\cal L}^{CM}(z) - \lambda\right) = 0
\ee
covers $N$ times elliptic curve (\ref{ell1}) with canonical holomorphic
1-differential
\be
dz = \frac{dx}{2y}
\ee
The SW BPS masses ${\bf a}$ and ${\bf a}_D$ are now the periods of the
generating 1-differential
\be\label{dSCal}
dS^{CM} = 2\lambda dz = \lambda{dx\over y}
\ee
along the non-contractable contours on $\Sigma^{CM}$
\footnote{Let us point out that
the curve (\ref{fscCal}) has genus $g = N$ (while in general the genus
of the curve defined by $N\times N$ matrix grows as $N^2$). However,
integrable system is still $2(N - 1)$-dimensional since the sum of the periods
of (\ref{dSCal}) vanishes due to specific properties of $\lambda dz$
and there are only $N - 1 = g - 1$ independent integrals of
motion.}.

In order to recover the Toda-chain system, one has to take the double-scaling
limit \cite{Ino}, when $m$ and $-i\tau$ both go
to infinity and
\be
q_i-q_j\rightarrow {\2}\left[(i-j)\log m +(q_i-q_j)\right]
\ee
so that the dimensionless coupling $\tau$ gets
substituted by a dimensionful parameter $\Lambda^{N} \sim
m^{N}e^{i\pi\tau}$. The idea is to separate the pairwise interacting particles
far away from each other and to adjust the coupling constant simultaneously
in such a way, that the interaction only of neighboring particles survives (and
turns to be exponential). In this limit, the elliptic curve degenerates into
a cylinder with coordinate $w = e^{z }e^{i\pi\tau}$ so that
\be
dS^{CM} \rightarrow dS^{TC} = \lambda\frac{dw}{w}
\ee
The Lax operator of the Calogero system turns into that of the
$N$-periodic Toda chain (\ref{LaxTC}):
\be
{\cal L}^{CM}(z )dz \rightarrow {\cal L}^{TC}(w)\frac{dw}{w}
\ee
and the spectral curve acquires the form (\ref{SpeC}).
In the simplest example of $N=2$
the spectral curve $\Sigma^{CM}$ has genus 2. Indeed,
in this particular case, Eq.~(\ref{fscCal}) turns into
\be\label{caln2}
{\cal P}(\lambda;x,y) = \lambda ^2 - u + m^2 x = 0
\ee
This equation says that with any value of $x$ one
associates two points of $\Sigma^{CM}$
\be
\lambda = \pm\sqrt{u - m^2 x}
\ee
i.e. it describes $\Sigma^{CM}$ as a
double covering of elliptic curve ramified at the points $x =
{u\over m^2}$ and $x = \infty$. In fact, $x = {u\over m^2}$ corresponds to a
{\it pair} of points on $E(\tau)$ distinguished by the sign of $y$. This
would be true for $x = \infty$ as well, but $x = \infty$ is one of the
branch points, thus, {\it
two} cuts between $x = {u\over m^2}$ and $x=\infty$ on every sheet of
$E(\tau)$ touching at the common end at $x=\infty$ become effectively a {\it
single} cut between $\left( {u\over m^2}, +\right)$ and $\left( {u\over
m^2}, -\right)$. Therefore, we can consider the spectral curve $\Sigma^{CM}$
as two tori $E(\tau)$ glued along one cut, i.e. $\Sigma^{CM}_{N=2}$ has
genus 2. It turns out to be a hyperelliptic curve (for $N = 2$ only!) after
substituting in (\ref{caln2}) $x$ from the second equation to the first one.

Two holomorphic 1-differentials on $\Sigma^{CM}$ ($g = N = 2$) can be
chosen to be
\be\label{holn2}
dv_+ = dz = \frac{dx}{2y} = \frac{\lambda d\lambda}{y}
\ \ \ \ \ \ \
dv_- = {dz\over\lambda} = \frac{dx}{2y\lambda}=\frac{d\lambda }{y}
\ee
so that
\be
dS = 2\lambda dz = \lambda{dx\over y} =
\frac{dx}{y}\sqrt{u - m^2 x}
\ee
and
\be
\frac{\partial dS}{\partial u} \cong \frac{dx}{2y\lambda} = dv_-
\label{CdSd}
\ee
The fact that only one of two holomorphic 1-differentials (\ref{holn2})
appears at the r.h.s. of (\ref{CdSd}) is related to their different parity
with respect to the ${\bf Z}_2\otimes {\bf Z}_2$ symmetry of $\Sigma^{CM}$:
$y \rightarrow -y$, $\lambda \rightarrow -\lambda$ and
$dv_{\pm}\rightarrow \pm dv_{\pm}$.
Since $dS$ has certain (positive) parity, its
integrals along the two of four elementary non-contractable cycles
on $\Sigma^{CM}$ automatically vanish leaving only two
non-vanishing quantities $a$ and $a_D$, as necessary for
the $4d$ interpretation.
Moreover, two rest nonzero periods can be defined
in terms of the "reduced" curve of genus $g=1$
\be
Y^2 = (y\lambda)^2 = \left(u - m^2 x\right)
\prod_{a=1}^3 (x - e_a),
\label{calN2red}
\ee
equipped with $dS = \left(u - m^2 x\right)\frac{dx}{Y}$
\footnote{This curve can be obtained by simple integration of the equations of
motion since we again here deal with the only degree of freedom in our integrable
system. The conserving energy $u = p^2 + m^2\wp (q)$ gives
\be
t = \int {dq\over\sqrt{h - m^2\wp (q )}} = \int {dx\over
\sqrt{(x-e_1)(x-e_2)(x-e_3)(u - m^2x)}}
\ee
i.e. exactly the Abel map on the reduced curve (\ref{calN2red})}.
Since for this curve $x = \infty$ is no more a ramification point, $dS$
has simple poles when $x = \infty$ (on both sheets of
$\Sigma^{CM}_{\rm reduced}$) with the residues $ \pm m$.

The opposite limit of the Calogero-Moser system
with vanishing coupling
constant $m^2 \rightarrow 0$ corresponds to the ${\cal N}=4$ SUSY
Yang-Mills theory with identically vanishing $\beta $-function.
The corresponding integrable
system is a collection of {\it free} particles and the generating differential
$dS = \sqrt{u}\cdot dz $ is just a {\it holomorphic} differential on
$E(\tau )$.

\subsection{Relativistic Toda Chain and the Ruijsenaars-Schneider Model}

So far we have considered the theories with only the Higgs and adjoint matter
contribution to the spectrum (\ref{spectrum}). If, however, one adds also the
soft KK modes \cite{Ne96}, the resulting integrable systems would correspond
to the Ruijsenaars-Schneider family \cite{RS,R,relToda}, which is often called
as the family of {\em relativistic} integrable models.
The 1-loop contributions to the effective charge (\ref{effcharge}) are now of
the form
\be\label{relKK}
T_{ij} \sim \sum _m\log
\left(a_{ij} + {m\over R_5}\right) \sim \log\prod _m\left(R_5a_{ij} + m\right)
\sim\log\sinh\left( R_5a_{ij}\right)
\ee
i.e. coming from 4D up to 5D one should make a trigonometric substitution
$a \rightarrow \sinh aR_5$, at least, in the formulas for the perturbative
prepotential. The similar change of variables corresponds to relativization
of integrable systems, which implies a sort of "Lie group
generalization" of the ``ordinary'' integrable systems related rather to
Lie algebras. The relativization of
an integrable system replaces the momenta of particles by their exponentials
and it results in the period matrices or effective charges of the
(\ref{relKK}) form. For example, in the case of $SU(2)$ pure gauge
theory it gives instead of Hamiltonian of the Toda chain (\ref{su2})
\be
\cosh z = \cosh p - {\tilde h}
\label{relsu2}
\ee
Generally, the net result is that in the case of relativistic Toda chain
the spectral curve is a minor modification of (\ref{specTC})
\be\label{specRTC}
w + {1\over w} = \left(\Lambda_5\lambda\right)^{-N/2}P(\lambda ),
\ee
which can be again rewritten as a {\em hyperelliptic} curve in terms of the
new variable $Y\equiv \left(\Lambda_5\lambda\right)^{N/2}\left(w-{1\over w}
\right)$
\be
Y^2 = P^2(\lambda ) - 4\Lambda_5^{2N}\lambda^{N}
\ee
where $\lambda \equiv \mu^2 \equiv e^{2\xi}$, $\xi $ can be chosen as
a spectral parameter of the relativistic Toda chain and $\Lambda_5$ is its
coupling constant.

The most general picture in these terms corresponds to the "unifying"
elliptic Ruijsenaars-Schneider model \cite{Ne96,BMMM2}.
The Lax operator for the elliptic Ruijsenaars model is\cite{R}
\be
{\cal L}^R_{ij} = e^{P_i} \frac{\sigma (q_{ij}+z)\sigma
(\epsilon)} {\sigma(q_{ij}+\epsilon)\sigma(z)} \nn \\ e^{P_i} = e^{p_i}
\prod_{k\neq i} \sqrt{\wp(\epsilon) - \wp (q_{ik})},
\label{RL}
\ee
and in the trigonometric limit it turns into
\be
{\cal L}^{TR}_{ij} =
e^{P_i} \frac{\sinh (q_{ij}+z)\sinh (\epsilon)}
{\sinh(q_{ij}+\epsilon)\sinh(z)}
\nn \\
e^{P_i} = e^{p_i}
\prod_{k\neq i} \sqrt{1 - {\sinh^2\epsilon\over\sinh^2(q_{ik})}}
\label{TRL}
\ee
Introducing $\nu_i = e^{2q_i}$, $\zeta = e^{2z}$ and $q = e^{2\epsilon}$ one
finds that
\be
{\cal L}^{TR}_{ij} =
e^{P_i} \frac{\zeta\nu_i - \nu_j}
{q\nu_i - \nu_j}{1-q\over 1-\zeta} \stackreb{\zeta\to\infty}{=}
(q-1){e^{P_i}\nu_i\over q\nu_i - \nu_j} + {\cal O}\left({1\over\zeta}\right)
\nn \\
e^{P_i} = e^{p_i}
\prod_{k\neq i} {\sqrt{q\nu_i-\nu_j}\sqrt{\nu_i-q\nu_j}
\over \nu_i - \nu_j}
\label{TRL1}
\ee
Only the leading term in (\ref{TRL1}) is often taken as an expression for
the Lax operator of trigonometric Ruijsenaars system.
The elliptic Ruijsenaars spectral curve acquires the form \cite{R,BMMM2}
\be
\det (\lambda - {\cal L}^R(z)) =
\sum_{k=0}^N (-\lambda)^{N-k}\left\{
\sum_{I_k} e^{P_{i_1}+\ldots+P_{i_k}}\tilde D_{I_k}\right\} =
\sum_{k=0}^N (-\lambda)^{N-k} D_k(z|\epsilon)\  H_k
=0
\label{Rsc}
\ee
where
\footnote{In \cite{BMMM2} we have used for technical reasons a slightly
different form of the Ruijsenaars spectral curve, which can be obtained from
Eq.~(\ref{Rsc}) substituting $\lambda\to\lambda c(z)$ (or $D_k(z|\epsilon)\to
D_k(z|\epsilon)c^k(z)$) with some (independent of Hamiltonians and
$\epsilon$) function $c(z)$. Such substitution does not change the "symplectic
form" ${d\lambda\over\lambda}\wedge dz$ and playing with such function some
particular formulas can be brought to more simple form.}
\be
D_k(z|\epsilon) =
(-)^{k(k-1)\over 2}
\frac{\sigma^{k-1}(z - \epsilon) \sigma(z + (k-1)\epsilon)}
{\sigma^k(z) \sigma^{k(k-1)}(\epsilon)}
\label{Dfun}
\ee
($D_0(z|\epsilon)=1$ and $D_1(z|\epsilon)=1$), and the generating
differential is
\be
dS \cong \log\lambda dz
\ee
In the simplest case of 1 degree of freedom ($N=2$) Eq.~(\ref{Rsc}) reads
\be
\lambda^2 - u\lambda + D_2(z|\epsilon) = \lambda^2 - u\lambda -
{\sigma(z-\epsilon)\sigma(z+\epsilon)\over\sigma^2(z)\sigma^2(\epsilon)} =
\lambda^2 - u\lambda + \wp(z) - \wp(\epsilon) = 0
\ee
and one gets a spectral curve identical to (\ref{caln2}). The same sort of
arguments as for the elliptic Calogero-Moser model shows that its full genus
is $g=2$ while the $SL(2)$ integrable system lives effectively on reduced
spectral curve of genus one. In trigonometric limit (\ref{TRL}), (\ref{TRL1})
the spectral curve (\ref{Rsc}) turns into (\ref{triRu}). An interested
reader can find the details concerning various limits of the
Ruijsenaars-Schneider model in \cite{BMMM2}.

\subsection{Fundamental Matter}

Let us turn, finally, to the case of SUSY QCD -- the SW theories with matter
multiplets in the lowest dimensional representations of the gauge group.
According to \cite{SW2,fumat} the spectral curves for \N2 SQCD with gauge
group $SU(N)$ and any number of matter multiplets $N_f < 2N$ have the form
\be
y^2 = P_N^2(\lambda ) - 4\Lambda^{2N-N_f}P_{N_f}(\lambda )
\\
P_{N}(\lambda) \equiv P^{(0)}_{N}(\lambda) +
R_{N-1}(\lambda) \equiv \prod _{i=1}^{N}(\lambda - \lambda_i(\bPhi,m))
\\
P_{N_f}(\lambda ) \equiv \prod_{\alpha = 1}^{N_f}
(\lambda - m_{\alpha})
\label{sqcd}
\ee
where $P_{N_f}(\lambda)$ and $R_{N-1}(\lambda)$ are "moduli independent"
polynomials of $\lambda$ -- i.e. they depend only on ``external''
masses of matter hypermultiplets \cite{fumat}, but spectral curves
are still described by hyperelliptic equation. In contrast to pure
gauge theory one runs immediately into a problem of parameter counting:
$N-1$ moduli ${1\over k}\Tr\bPhi^k$ together with $N_f$ masses $m_{\alpha}$
give $N+N_f-1$ and this is {\em more} than hyperelliptic moduli $2g-2=2N-4$
of the equation (\ref{sqcd}) for $N_f>N-3$. In other words, the moduli space
of eq.~(\ref{sqcd}) is too small to be parameterized by {\em all} parameters
of the theory, that is why there are some complications in retranslating
these models into the language of integrable systems
\footnote{In particular, there are different ideas of interpretation of
the equation (\ref{sqcd}) (at least in the case of small $N_f<N$) in the
context of integrable models, some of them can be found in
\cite{M4,AhN,KriPho}. We shall consider below, following \cite{XXX,Spin2}
only one of such options, which seems, however, to be the most attractive
at present since it leads to a family of integrable systems which admits,
for example, as in the case of adjoint matter, inclusion of the KK sector, or
natural relativistic generalizations.}. However, the form (\ref{sqcd})
immediately implies \cite{M4} that for
\N2 SQCD there should exist a $2\times 2$ representation of the kind
(\ref{specTC0}) with monodromy matrix $T_N(\lambda )$, whose invariants have
the form
\be\label{trdet}
\Tr\ T_{N}(\lambda) = P_{N}(\lambda)
\\
\det T_{N}(\lambda ) = P_{N_f}(\lambda )
\ee
A natural proposal then is \cite{XXX} to use a generalization of the
$2\times 2$ Toda chain Lax representation, i.e. to deform the Lax matrix
(\ref{LTC}) {\em preserving} the Poisson brackets (\ref{quadrP}) and
(\ref{quadrT}).
The monodromy matrix $T(\lambda)$ is again constructed by multiplication
of $L_i(\lambda)$'s (satisfying (\ref{quadrP})) at different sites
giving rise to the spectral curve equation
\be\label{fsc-SCh}
\det\left(T_{N}(\lambda) -  w\right) = 0
\ee
where from $T$-matrix one has to require (\ref{trdet}). Wide class of such
systems is given by the family of {\em spin chains} and to get most general
form of
the transfer matrix one should consider the {\em inhomogeneous} spin chain
\footnote{For the Toda chain the inhomogenity parameters $l_i$ are
not independent variables -- they can be reabsorbed into the definition of
momenta.}
\be\label{t-mat}
T_{N}(\lambda ) = \prod_{N \ge i\ge 1}^{\curvearrowleft} L_i(\lambda
-l_i )
\ee
If Lax matrices $L_i(\lambda )$ were of the size $p\times p$ the equation
(\ref{fsc-SCh}) would have the form of $p$-th degree polynomial in $w$,
if $p=2$ it is exactly the general case of (\ref{sqcd})
\footnote{We introduce new variable $W={w\over\sqrt{\det T_N(\lambda)}}$ in
order to make analytic representation of spectral curve more symmetric. In the
picture of M-theory these redefinitions are related with different brane
pictures of the theories where presence of fundamental matter multiplets
are caused by extra semi-infinite branes.}
\be\label{fsc-sc1}
w + \frac{\det T_{N}(\lambda)}{w} =
 \Tr T_{N}(\lambda ),
\ee
or
\be
W + \frac{1}{W} = \frac{\Tr T_{N}(\lambda)}
{\sqrt{\det T_{N}(\lambda)}} \equiv {P_{N}(\lambda )\over
\sqrt{P_{N_f}(\lambda )}}
\label{fsc-sc2}
\ee
when $\Tr T_N(\lambda )$ has, like in the Toda chain case, degree $N$ but
$\det T_N(\lambda )$ is not a unity, but some general polynomial of
degree $N_f\leq 2N$.
The generating 1-form, according to general rules, becomes
\be\label{1f}
dS = \lambda\frac{dW}{W}
\ee
The r.h.s. of the equations (\ref{fsc-sc1}), (\ref{fsc-sc2}) contain the
dynamical variables of
the spin system only in special combinations -- the
Hamiltonians (which are all in involution, i.e. mutually Poisson-commuting
or the Casimir functions and inhomogenities, commuting with all dynamical
variables). The $2\times 2$ Lax
matrix for the simplest $sl(2)$ {\em rational} $XXX$ chain is
\be\label{laxmatr}
L(\lambda) = \lambda \cdot {\bf 1} + \sum_{a=1}^3 S_a\cdot\sigma^a.
\ee
and the Poisson brackets on the space of the dynamical variables $S_a$,
$a=1,2,3$ are implied by {\em quadratic} r-matrix relations (\ref{quadrP})
with the same rational $r$-matrix, as for the Toda chain \cite{Skl}.

The $r$-matrix relations (\ref{quadrP}) for the Lax matrix (\ref{laxmatr}) are
equivalent to well-known $sl(2)$ commutation (Poisson bracket) relations
\be\label{Scomrel}
\{S_a,S_b\} = i\epsilon_{abc} S_c,
\ee
i.e. vector $\{S_a\}$ plays the role of angular momentum (``classical spin'')
variables giving the name ``spin-chains'' to the whole class of systems.
The algebra (\ref{Scomrel}) has an obvious Casimir operator
(an invariant Poisson commuting with all generators $S_a$),
\be\label{Cas}
K^2 = {\bf S}^2 = \sum_{a=1}^3 S_aS_a,
\ee
so that
\be\label{detTxxx}
\det_{2\times 2} L(\lambda) = \lambda^2 - K^2,
\nn \\
\det_{2\times 2} T_{N}(\lambda) = \prod_{1\le i\le N}
\det_{2\times 2} L_i(\lambda-l_i) =
\prod_{1\le i\le N}\left((\lambda - l_i)^2 - K_i^2\right) = \nn \\
= \prod_{1\le i\le N}(\lambda + m_i^+)(\lambda + m_i^-)
\ee
where we assumed that the values of spin $K$ can be different at
different nodes of the chain, and
\footnote{Eq.~(\ref{mpm}) implies that the limit of vanishing masses, all
$m_i^\pm = 0$, is associated with the {\it homogeneous} chain
(all $l_i = 0$) and vanishing spins at each site (all $K_i = 0$).}
\be
m_i^{\pm} = -l_i \mp K_i.
\label{mpm}
\ee
Determinant of monodromy matrix (\ref{detTxxx})
depends on dynamical variables only through the Casimirs $K_i$ of the
Poisson algebra, and trace $P_{N}(\lambda) =
\frac{1}{2}\Tr_{2\times 2}T_{N}(\lambda)$
generates the Hamiltonians or integrals of motion.
In the case of $sl(2)$ $XXX$ spin chains, the spectral
equation acquires exactly the form
(\ref{fsc-sc1}) or (\ref{fsc-sc2}) where the number of matter multiplets,
which determines the degree of polynomials in (\ref{trdet}),
$N_f\le 2N$ depends on a particular ``degeneracy'' of full chain.

In this picture the rational $sl(2)$ $XXX$ spin chain literally corresponds to
a $N_f < 2N$ \N2 SUSY QCD. Things are not so simple with the
``conformal'' $N_f = 2N$ case when an additional dimensionless parameter
appears -- a naive toric generalization of the $XXX$-picture leads to the
$XYZ$ chain with the Hamiltonian structure given by elliptic
Sklyanin algebra \cite{Skl}. This model was proposed in \cite{Spin2}
as a candidate for integrable system behind the $N_f = 2N$ theory,
and, as it was discussed later (see, for example, \cite{mami97} and
references therein), it should be rather interpreted as a 4D theory with
two extra compact dimensions (or a compactified 6D theory)
\footnote{Indeed, in 6D theory with {\em two} extra compactified dimensions of radii
$R_5$ and $R_6$ one should naively expect
\be\label{6dKK}
T_{ij} \sim \sum _{m,n}\log
\left(a_{ij} +
{m\over R_5} +{n\over R_6}
\right) \sim
\\
\sim\log\prod _{m,n}\left(R_5a_{ij} + m+n{R_5\over R_6}\right)
\sim\log\theta\left(R_5a_{ij}\left|i{R_5\over R_6}\right)\right.
\ee
i.e. coming from 4D or 5D to $D=6$ one should replace the rational
(trigonometric) expressions by the elliptic functions, at least, in the
formulas for the perturbative prepotential, the (imaginary part of) modular
parameter being identified with the ratio of the compactification
radii $R_5/R_6$.}.
The condition $N_f=2N$ which can be easily broken in 4D
and 5D situations is very strict in 6D and one may think of the corresponding
theory as of blowing up all possible compactified
dimensions in the $N_f=2N$ 4D theory with vanishing $\beta$-function.

In general, one finds, that the spectral curves for SUSY QCD with $N_f$
fundamental multiplets (and prepotentials) in 4D, 5D and 6D cases are
described by similar formulas \cite{mami97}.
The spectral curves in all cases can be written in the form
\be\label{scg}
w+{Q^{(d)}(\xi)\over w}=2P^{(d)}(\xi)
\ee
or
\be\label{scg'}
W+{1\over W}={2P^{(d)}(\xi)\over\sqrt{Q^{(d)}(\xi)}},
\ \ \ \ \ \ W = {w\over\sqrt{Q^{(d)}(\xi)}}
\ee
the generating differentials are
\be
dS=\xi d\log W
\ee
and the perturbative part of the prepotential is always of the form
\be
{\cal F}={1\over 4}\sum_{i,j}f^{(d)}(a_{ij})-{1\over 4}\sum_{i,\alpha}
f^{(d)}(a_i-m_{\alpha})+{1\over 16}\sum_{\alpha,\beta}
f^{(d)}(m_{\alpha}-m_{\beta})
\label{pertd}
\ee
The functions introduced $f^{(d)}$ are:
\be
Q^{(4)}(\xi)\sim\prod_\alpha^{N_f}(\xi-m_\alpha),\ \ \
Q^{(5)}(\xi)\sim\prod_\alpha^{N_f}\sinh(\xi-m_\alpha)\\
Q^{(6)}(\xi)\sim\prod_\alpha^{N_f}{\theta_*(\xi-m_\alpha)\over\theta_*^2
(\xi-\xi_i)}
\ee
\be
P^{(4)}\sim\prod_i^{N}(\xi-a_i),\ \ \
P^{(5)}\sim\prod_i^{N}\sinh(\xi-a_i)\\
P^{(6)}\sim\prod_i^{N}{\theta_*(\xi-a_i)\over\theta_*(\xi-\xi_i)}
\ee
(in $P^{(5)}(\xi)$, there is also some exponential of $\xi$ unless $N_f=2N$)
\be
f^{(4)}(x)=x^2\log x\\
f^{(5)}(x)=\sum_{n}f^{(4)}\left(x+{n\over R_5}\right)=
{1\over 3}\left|x^3
\right|-{1\over 2}{\rm Li}_3\left(e^{-2|x|}\right)\\
f^{(6)}(x)=\sum_{m,n}f^{(4)}\left(x+\frac{n}{R_{5}}+\frac{m}{R_{6}}\right)=
\sum_n f^{(5)}\left(x+n{R_{5}\over R_6}\right)=
\\
= \left({1
\over 3}\left|x^3\right|
-{1\over 2}{\rm Li}_{3,q}\left(e^{-2|x|}\right)
+ {\rm quadratic\ \ terms}\right)
\label{fung}
\ee
so that
\be
{f^{(4)}}''=\log x\ \ \ {f^{(5)}}''(x)=\log\sinh x\ \ \
{f^{(6)}}''(x)=\log\theta_*(x)
\label{fpertd}
\ee
Note that, in 6D case, we have always $N_f = 2N$. The variables
$\xi_i$ above are inhomogenities in integrable system, and, in $D=5,6$,
there is a restriction $\sum a_i=\sum\xi_i=\2\sum m_\alpha$
which implies that gauge moduli would be
rather associated with $a_i$ shifted by the constant
${1\over 2N}\sum m_\alpha$.

\section{Conclusion}

In this talk I have tried to formulate main issues of the correspondence
between the Seiberg-Witten curves governing the exact solutions to \N2 SUSY
gauge theories and integrable systems. During last 4 years there was a lot
of progress in this direction. However, there is still a lot of questions which
are not yet understood and deserve further investigation. Let me, finally,
point out at least some of them:

\begin{itemize}
\item The first, and the most essential question is still open: how to
derive the SW -- integrable systems correspondence from first principles.
There is no clear answer to the question what means the dynamics governed by
an integrable system directly in terms of SUSY gauge theory.
\item The point which seems to be already clear is that SW construction
itself becomes much more transparent from the point of view of
non-perturbative string theory or M-theory. However, it is still many open
questions in this direction -- for example the M-theory picture of the
Ruijsenaars-Schneider model, double-elliptic systems etc. It is clear that
the Ruijsenaars Lax operator (\ref{RL}) satisfies similar to (\ref{gauss})
$\bar{\partial}$-equation, but there is no clear interpretation of this
equation in terms of D-brane constructions.
\item I have already mentioned some general problems with fundamental
matter. Of particular interest is the conformal $N_f = 2N$ case, which
should be clearly related, on one hand, with double-elliptic systems and, on
the other hand, with the K3 and elliptic Calabi-Yau compactifications of
string and M-theory.
\item In this talk only the $SU(N)$ gauge theories were considered in
detail. I did not touch at all many problems related to the generalizations
to the other gauge groups and representations. From the point of view of
integrable systems this is a question, in part, about different
representations of the Lax pairs for integrable systems and it is overlapped,
thus, with the problem of different Lax pairs for the Calogero-Moser systems
found recently in \cite{BCS} and \cite{DPh}.

\end{itemize}

\section{Acknowledgements}

I am grateful to H.Braden, I.Krichever, A.Mironov and A.Morozov for
illuminating discussions and to the organizers of the conference for
nice time in Edinburgh. The work was partially supported by the RFBR
grant 98-01-00344 and the INTAS grant 96-482.

\end{document}